\documentclass[aps,prx,twocolumn,amsmath,amssymb,amsfonts,nofootinbib,superscriptaddress,altaffilletter]{revtex4-2}

\usepackage{graphicx}
\usepackage{hyperref}
\hypersetup{
  bookmarksopen=true
}
\usepackage{ulem}
\usepackage{amssymb}
\usepackage{amsmath}
\usepackage{color}
\usepackage{supertabular}
\usepackage{dcolumn}
\usepackage[mathscr]{eucal}
\usepackage{mathrsfs}

\def\sci#1#2{#1\times10^{#2}}

\def\RAJ{\textrm{RA}_{\textrm J2000}}
\def\DECJ{\textrm{DEC}_{\textrm J2000}}
\def\mathL{\mathcal L}

\begin{document}

\title{A frequency-resolved atlas of the sky in continuous gravitational waves}

\author{Vladimir Dergachev}
\email{vladimir.dergachev@aei.mpg.de}
\affiliation{Max Planck Institute for Gravitational Physics (Albert Einstein Institute), Callinstrasse 38, 30167 Hannover, Germany}
\affiliation{Leibniz Universit\"at Hannover, D-30167 Hannover, Germany}

\author{Maria Alessandra Papa}
\email{maria.alessandra.papa@aei.mpg.de}
\affiliation{Max Planck Institute for Gravitational Physics (Albert Einstein Institute), Callinstrasse 38, 30167 Hannover, Germany}
\affiliation{Leibniz Universit\"at Hannover, D-30167 Hannover, Germany}
\affiliation{University of Wisconsin Milwaukee, 3135 N Maryland Ave, Milwaukee, WI 53211, USA}

\begin{abstract}
We present the first atlas of the continuous gravitational wave sky, produced using LIGO O3a public data. For each $0.045$\,Hz frequency band and every point on the sky the atlas provides gravitational wave amplitude upper limits, signal-to-noise ratios (SNR) and frequencies where the search measures the maximum SNR. The approximately top $1.5$\% of the atlas results are reanalyzed with a series of more sensitive searches with the purpose of finding high SNR long coherence signals from isolated neutron stars. However, these searches do not reveal the presence of such signals. The results presented in the atlas are produced with the Falcon pipeline and cover nearly monochromatic gravitational-wave signals in the 500-1000\,Hz band, with up to $\pm \sci{5}{-11}$\,Hz/s frequency derivative. 
The Falcon pipeline computes loosely coherent power estimates to search data using a succession of coherence lengths. For this search we used 6 months of data, started with a 12 hour coherence length and progressed to 6 days. Compared to the most sensitive results previously published (also produced with the Falcon pipeline) our upper limits are 50\% more constraining. Neutron stars with ellipticity of $10^{-8}$ can be detected up to 150\,pc away, while allowing for a large fraction of the stars' energy to be lost through non-gravitational channels. These results are within an order of magnitude of the {\it minimum} neutron star ellipticity of $10^{-9}$ suggested in \cite{ellipticity}.
\end{abstract}

\maketitle

\section{Introduction}

Astronomy begun with the observation of light from persistent sources. Gravitational-wave astronomy, on the contrary, started with the observation of signals from the merger of compact stars, which are loud transient events lasting less than a minute \cite{LIGOScientific:2021djp}.  

This paper is about a different type of gravitational wave signals: those consistently ``ON" over months and even years. Such signals, commonly referred to as continuous gravitational waves, are expected from a variety of astrophysical scenarios and could be detected by the LIGO, Virgo or KAGRA interferometers \cite{aligo, avirgo, KAGRA:2022fgc}. Among the scenarios that could give rise to continuous waves detectable by ground-based detectors  there are rotating neutron stars emitting due to a sustained quadrupolar deformation, fluid oscillations \cite{Haskell:2015psa,Lindblom:1998wf, bo_rmodes} as well as more exotic scenarios as superradiance from ultralight bosons around spinning black holes \cite{superradiance, Zhu:2020tht,LIGOScientific:2021jlr}.

The detection of a continuous gravitational wave will lead to high-precision tests of General Relativity and, depending on the emission mechanism, may probe the interior and physics of neutron stars in an entirely new way or unveil new physics \cite{Arvanitaki:2009fg}. For this reason there exist a variety of research efforts aiming at detecting such signals \cite{keith_review}. 

Computational searches combining data collected over weeks or even longer periods
are carried out using a broad variety of approaches. This variety arises because the most sensitive methods cannot be used due to computational limitations so different speed-optimized methods have been designed, that sacrifice either breadth of search or sensitivity, or a bit of both.  Our loosely coherent algorithms \cite{loosely_coherent, loosely_coherent2, loosely_coherent3} and, in particular, the Falcon search pipeline \cite{O2_falcon, O2_falcon2, O2_falcon3} that is used for the analysis presented here, have performed very well in this trade-off. 

In this paper we provide a window on the universe of continuous gravitational wave sources. For the first time ever, every sky location has associated a spectrum of upper limits on the gravitational wave amplitude, as well as a signal-to-noise ratio (SNR) spectrum. These are, nearly-raw, minimally processed results and access to them will enable new and independent searches, as we explain in Section \ref{sec:results}. Together with the new Einstein@Home results \cite{EatHO3a}, these are the most sensitive results to date. 

As it is common in broad surveys for continuous gravitational waves, we follow-up and investigate less than 2\% of the results of our search -- and such results are marked in the atlas. None of the followups results in a conclusive association with astrophysical sources. The rest of the atlas remains unexplored. Other scientists can use the atlas in combination with electromagnetic and particle data \cite{JST, CTA, SKA, AMS, PTF}, and catalogs \cite{SIMBAD, ATNF}, in search of significant associations. Many other studies are possible. In the appendix, we provide an example of an interesting search that can be easily performed using only the atlas data.

The atlas contains nearly two billion records. To make it easily accessible, the atlas has been designed so that it can be analyzed on a small personal computer using the MVL library \cite{RMVL, MVL}.  We provide open-source software (compatible with Linux, Windows and Mac OS) that produces sky maps for wide frequency bands, allows queries of the atlas by sky position and/or frequency, and performs computations using the full scan of the data. Effectively this enables searches within the covered parameter space with the full sensitivity of the Falcon pipeline, but without the need for a large computing cluster. Our hope is to make gravitational wave studies and discoveries accessible to scientists with a modest computing budget.

In the following sections we describe the gravitational wave signals, 
the analysis method, the search setup (including the follow-up) and the results of the analysis. 

\section{Continuous gravitational wave sources}
The ground-based gravitational wave detectors presently in operation \cite{aligo, avirgo,KAGRA:2022fgc} are sensitive in the audio frequency range up to a few kHz, with the lower bound at a few tens of Hz given by the steep ``seismic wall''.

The seismic wall at low frequencies precludes the observation of continuous gravitational waves from the decay of the orbit of white dwarf binaries with large separation. For this reason the classical sources of audio-frequency continuous gravitational wave signals are rotating neutron stars with non-axisymmetric deformations. For such sources the gravitationl wave emission is powered by the stars' rotation. 

Theory predicts that the neutron star crust can support quadrupolar deformations of $10^{-6}$ and perhaps even larger for neutron stars with exotic composition, however signals like this have not yet been detected \cite{Johnson-McDaniel:2012wbj,Gittins:2020cvx,Gittins:2021zpv}. 

\begin{figure}[htbp]
\includegraphics[width=3.3in]{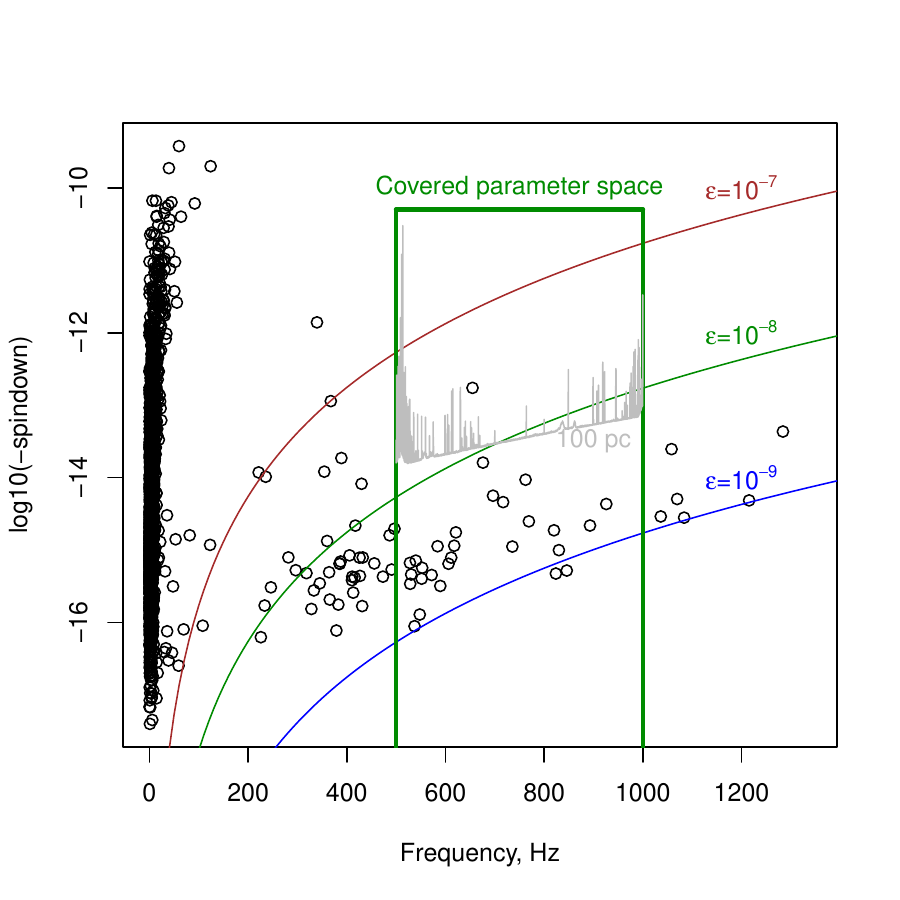}
\caption[Pulsar parameter space]{
\label{fig:pulsars}
This plot shows the frequency and frequency-derivative of isolated neutron stars from the ATNF catalog \cite{ATNF} (circles). The pulsar frequency is shown multiplied by a factor of 2, to correspond to the expected gravitational wave frequency. The curves show the frequency/frequency-derivative combinations for a gravitar (a neutron star loosing energy solely due to gravitational wave emission) for different values of the ellipticity. Following \cite{ellipticity} we observe that known pulsars have spin-down values that lie at or above the curve $\epsilon=10^{-9}$. For gravitars this would imply that the minimum ellipticity is around $10^{-9}$. 
The region above the gray trace and enclosed by the box of green straight lines shows the frequency/frequency-derivative combinations of optimally oriented gravitars at 100\,pc, that our search could detect. The frequency-derivative of a detectable gravitar is proportional to the square of the distance, so if the distance increases by a factor of 10, correspondingly the grey trace moves up by a factor of 100.
}
\end{figure}

A constraint on the likehood of existence of rapidly spinning neutron stars with deformations much larger than $10^{-8}$ comes from the lifetime of sources: due to emission of the gravitational waves the spin rate of the neutron star decreases in time. Since the energy loss is proportional to the sixth power of the rotation rate, highly deformed (highly emitting) high frequency sources are less likely to exist in our galaxy because they quickly become slowly rotating neutron stars \cite{Pagliaro:2023bvi}. In fact, known radio pulsars rotating faster than $250$\,Hz ($500$\,Hz gravitational wave frequency) have frequency derivative less than $\sci{2}{-13}$\,Hz/s \cite{ATNF} (Figure \ref{fig:pulsars}), corresponding to maximum deformations of the order of $10^{-8}$ and below. 

Examination of  properties of known pulsars suggests that a population might exist whose spin evolution is governed by gravitational wave emission, with typical ellipticities in the range of $\approx 10^{-9}$ to $10^{-8}$ \cite{ellipticity} (see Figure~\ref{fig:pulsars}).

\section{The analysis method}

The expected signals are nearly monochromatic, with small frequency drifts in time. The frequency $f(t)$ of such signals is well described by Taylor polynomials with few components \cite{O2_falcon2}. For our search the second order Taylor polynomial is sufficient:

\begin{equation}
f(t)=f_0+(t-t_0)f_1+(t-t_0)^2f_2/2,
\label{eq:freqEvolution}
\end{equation}
We call these ``IT2'' type signals, where ``I'' indicates an isolated source (no binary modulation), ``T'' indicates Taylor polynomial model and ``2'' gives degree of the polynomial \cite{O2_falcon2}.

The signal at the detector is more complex as it presents frequency modulations due to the Doppler effect induced by the relative motion between the source and the receiver. For isolated neutron stars, which are the target of this paper, the largest Doppler shifts are due to Earth orbital motion \cite{Jaranowski:1998qm, Fourier}, and are on the order of $10^{-4}~f_0$, with $f_0$ being the signal frequency. Additionally the signal also presents amplitude modulations due to the non-uniform beam-pattern response of the detector across the sky. 

The first step of the analysis consists in heterodyning the data at a frequency close to the target signal frequencies and in extracting a narrow band from the heterodyned data. This step is repeated over the different frequency sub-bands which compound the broad search band and it is efficiently accomplished in the Fourier domain. 

After this transformation the data for the band of interest is a complex time series $\left\{z(t)\right\}$, down-sampled with respect to the original data. In this representation, the gravitational wave signal has the form
\begin{equation}
S(t)=h_0A(t) e^{i \Phi(t) },
\end{equation}
where $h_0$ is the intrinsic amplitude of the signal at the detector and the amplitude response $A(t)$ depends on the relative orientation of the source and the detector and hence changes as the Earth rotates. The phase evolution term $\Phi(t)$ at the detector is due to the Doppler-shifted  frequency evolution of Eq.~\ref{eq:freqEvolution}, it is independent of orientation but varies with sky location, frequency and frequency derivatives \cite{Fourier,Jaranowski:1998qm}. 

The input data is then the sum of the unknown signal and noise $n(t)$:
\begin{equation}
z(t)= h_0 A(t) e^{i \Phi(t) }+n(t)
\end{equation}
The term $n(t)$  is a combination of fundamental noise sources and ``technical" noise sources. The former include shot noise and radiation pressure on the instrument mirrors, and can be considered Gaussian. 
The technical noise sources are due to accidental couplings between noise sources external and internal to the detector and the gravitational wave output channels. These couplings are highly dependent on frequency and are non-stationary. Since a lot of effort has gone into reducing these couplings, there are many ``clean'' bands where these couplings are small enough to be practically negligible. Even in these clean bands, however, the $n(t)$ is in general not a stationary process -- the reason being that the beam power in the detector arms varies in time affecting both the level of fundamental noise sources and the gain of the detector. 

Next we consider data with the amplitude response and phase evolution terms taken out:
\begin{equation}
\label{eq:ztilde}
\tilde{z}(t)=\frac{z(t)e^{-i\Phi_0(t)}}{A(t)}=h_0 e^{ i (\Phi(t)-\Phi_0(t)) } + \frac{ n(t)e^{ -i\Phi_0(t)} } {A(t)} .
\end{equation}
$\Phi_0(t)$ is an assumed phase evolution, in general distinct from the true evolution $\Phi(t)$ when we search for unknown signals. We assume that the amplitude evolution for a signal with phase $\Phi_0(t)$ is practically equal to $A(t)$ of the signal with phase evolution $\Phi(t)$ which is correct for signals from sky locations within~$1^\circ$ of each other and the same gravitational wave polarization. A generic sky position can also be considered leading to the slightly different expression \cite{loosely_coherent2} actually used in the search, but here we make this simplifying assumption in order to illustrate the main idea.

We construct a linear time-domain filter ${\mathcal L}$, say a band-pass filter, with wide enough bandwidth to leave invariant signals of the type
\begin{equation}
\label{eq:DeltaPhi}
 h_0e^{i (\Phi(t)-\Phi_0(t)) }
\end{equation}
and with as narrow bandwidth as desired to reject the maximum possible amount of noise. 

The power $P=\left\| \mathL\tilde{z}\right\|^2$ can then be used to establish a limit on the strength $h_0$ of the gravitational wave signal in the data. This is best achieved based on excess power measurements, i.e. by considering a set of power measurements for a set of phase evolution models $\Phi_0$ and by setting an upper limit under the assumption that only a single signal is present \cite{universal_statistics}.

In our approach there is an inherent trade-off between the maximum phase mismatch of phase evolutions $\left\{\Phi(t)\right\}$ from $\Phi_0(t)$ and the narrowness of the filter $\mathL$. The narrower is the filter, the less noise it lets through and the more sensitive the search is.

The structure of the manifold of  phase evolutions $\left\{\Phi(t)\right\}$ used to design the filter is very important but its description goes well beyond the scope of this paper \cite{Fourier}. 

An important characteristic of the manifold of phase evolutions is the maximum phase variation $V$ over the timescale $\delta$:
\begin{equation}
V(\delta)=\max_{|t_1-t_2|<\delta}\left|\Phi(t_1)-\Phi_0(t_1)-\Phi(t_2)+\Phi_0(t_2)\right|.
\end{equation}
When $V(\delta)$ is close to $0$, the filter $\mathL$ can be a simple averaging of the data. As the variation grows, the filter $\mathL$ needs to include an explicit - and more computationally expensive - demodulation. 
As $V(\delta)$ exceeds $\pi$, the exponent $\exp i\left(\Phi(t)-\Phi_0(t)\right)$ changes sign. 
The value of $\delta$ when $V(\delta)=\pi$ is known as coherence length.

By picking $\mathL$ as projection onto a constant, we obtain a very narrow filter that admits only $\Phi(t)-\Phi_0(t)=\textrm{const}$ - this is known as a ``fully-coherent'' search. At the other extreme, if $\mathL$ is the identity operator which leaves its input unchanged, the power $P=\left\| \mathL\tilde{z}\right\|^2=\left\|\tilde{z}\right\|^2$ discards the phase information and we arrive at a method with the shortest coherence length possible given the bandwidth of the heterodyne procedure used to obtain $z(t)$. 
The loosely coherent method \cite{loosely_coherent, loosely_coherent2, loosely_coherent3} employed by Falcon uses the filter $\mathL$ with a tunable coherence length \cite{loosely_coherent3}.

In practice in order to search for unknown sources anywhere on the sky one needs to repeat the above procedure for many different assumed phase evolutions $\Phi_0$, corresponding to different source sky positions and signal frequency and frequency derivatives. 
We thus arrive at the fundamental difficulty of all-sky searches: the number of different phase evolutions that need to be considered and hence the computational demands, scales with coherence length, which refines the resolution in sky, frequency and frequency-derivatives.

The Falcon pipeline used in this paper employs a loosely coherent algorithm \cite{loosely_coherent, loosely_coherent2, loosely_coherent3} that finds the maximum power $P$ for a set of assumed evolutions $\left\{ \Phi_0^i(t)\right\}$. Operating on a set allows to frame the problem as the optimization of a highly oscillatory function and take advantage of efficiencies of scale. 

The typical analysis takes several months to complete on a facility of the scale of the ATLAS computing cluster \cite{ATLAS}, so with every new analysis we strive to better take advantage of existing data and computational infrastructure. While limits of computational hardware dictate the choice of the algorithm used, compilation of the code, scheduling, threading of the compute jobs on the physical cores and 
even aspects such as data layout are carefully considered for every run because they can have a large impact on computational speed.

Based on the power $P$, the algorithm establishes upper limits \cite{universal_statistics} and SNRs for every analyzed portion of the sky and every frequency band.
This data set is quite large and in the past was considered impractical to record, so only summary information was presented, obtained by grouping the results by frequency.

For this search, instead, thanks to the use of Mapped Vector Library (MVL) files, we could keep and distribute all the results -- upper limits and SNRs, adding up to over 70\,GB of data. The MVL technology was originally used to extract large amounts of engineering data describing the pipeline's operation in real time. Only later it was realized that it is also perfectly suited for the scientific output of the pipeline.  The key property of MVL files is that they are optimized for memory mapping. For example, with a notebook with 16 GB RAM and solid state drive, one can open the atlas in its entirety and access any data by simple array subscription, as if the notebook had an order of magnitude more memory than it actually does.

\section{The search}

Our search uses data of the LIGO H1 and L1 interferometers, from the publicly released O3a set \cite{O3aDataSet}. We limit our search to the LIGO detectors because no other detector comes within a factor of 2 of their sensitivity. We set O3b data \cite{O3bDataSet} aside as an independent data set for verification of any surviving outlier. This is a standard approach for searches over a very broad parameter space: due to the large trials' factor, noise alone produces outliers that can only be vetoed through verification on a different data set (see for example \cite{Papa:2020vfz}). We note that this procedure can be used because the same physical signal is assumed present in both data sets.

The O3a data set is contaminated by loud short-duration excesses of noise \cite{lvc_O3_allsky, O2O3_detchar, Steltner:2021qjy}. To address this, we first prepare short Hann-windowed Fourier transforms (SFTs) with 0.25\,s duration. We then monitor the frequency band 20-36\,Hz for power excess that indicates noise contamination and exclude the affected SFTs from the analysis. The remaining SFTs are searched with the Falcon pipeline. The 20-36\,Hz band was chosen because it is outside of the search range and heuristically we found it to be an effective  ``witness band" to identify spectral disturbances. 

We search for nearly monochromatic signals of the IT2 type given by Eq.~\ref{eq:freqEvolution}
with the signal frequency $f_0$ and its first derivative $f_1$ defined at GPS epoch $t_0=1246070000$ (2019 Jul 2 02:33:02 UTC).

The search targets emission in the $500-1000$\,Hz frequency range, covering frequency derivatives $|f_1|\le \sci{5}{-11}$\,Hz/s and $|f_2|\le \sci{4}{-20}$\,Hz/$\textrm{s}^2$. Compared to our earlier analyses on O2 data \cite{O2_falcon3} we use a more sensitive data set and we search a spindown range larger by a factor $\gtrsim$ 16. This provides a larger margin ($\times 2.9$ at $1000$\,Hz) for Doppler shifts due to stellar motion  and energy loss due to magnetic fields (Figure \ref{fig:pulsars}).

\begin{table}[htbp]
\begin{center}
\begin{tabular}{ccc}\hline
Stage & \multicolumn{1}{c}{Coherence length } & \multicolumn{1}{c}{SNR}\\
 & \multicolumn{1}{c}{(days)} & \multicolumn{1}{c}{threshold}\\
\hline
\hline
1  & 0.5 & 6 \\
2  & 1 & 8 \\
3  & 2 & 8 \\
4  & 6 & 12 \\
\hline
\end{tabular}
\end{center}
\caption{Parameters for each stage of the search. 
Stage 4 refines outlier parameters by using a denser sampling of parameter space, and then subjects them to an additional consistency check based on the analyses of individual interferometer data. The results from stages 1 and 2 are used to construct the atlas, stages 3 and 4 are solely outlier follow-ups.
}
\label{tab:pipeline_parameters}
\end{table}

The analyzed frequency band was chosen to make the first gravitational wave atlas simpler to use, while providing the most useful data. The lower cutoff of 500\,Hz was picked to avoid numerous detector artifacts, while the upper cutoff of 1000\,Hz was imposed to limit total data size, as the size of the atlas scales as cube of upper frequency.

The pipeline employs 4 stages (Table \ref{tab:pipeline_parameters}) with coherence length increasing from 12 hours to 6 days. Results exceeding the SNR threshold of one stage pass to the next stage. The upper limits and the atlas are constructed based on the results of the first two stages only. This simplifies the upper limit procedure without significant impact on the sensitivity. The last stage of the pipeline has long enough coherence time that signals that have survived up to that stage can now be identified using the data from the individual interferometers separately. This means that signal-consistency checks can be performed by comparing the results from each detector.

\section{Results}
\label{sec:results}

\begin{figure}[htbp]
\includegraphics[width=3.3in]{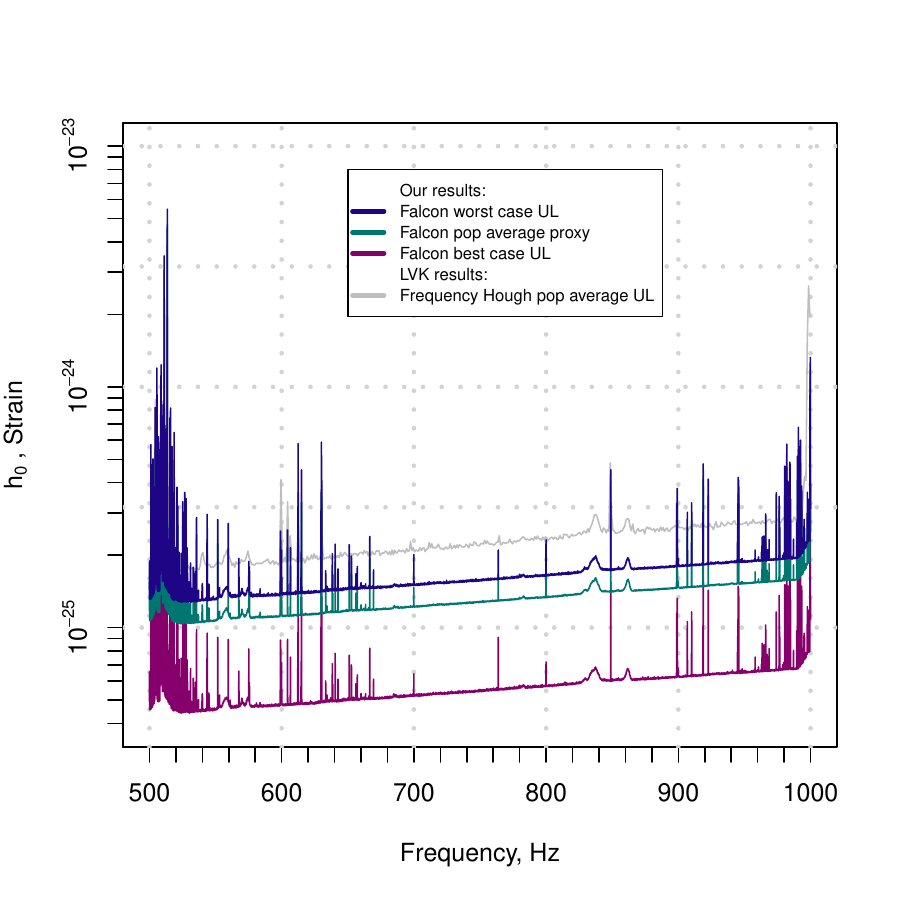}
\caption[Upper limits]{
\label{fig:amplitudeULs}
Gravitational wave intrinsic amplitude $h_0$ upper limits at 95\% confidence as a function of signal frequency. The upper limits are a measure of the sensitivity of the search. We introduce a ``population average" proxy upper limit in order to compare with the latest LIGO/Virgo all-sky results \cite{lvc_O3_allsky2}.  In this frequency range the results from \cite{lvc_O3_allsky2} are a factor $\gtrsim$ 1.65 less constraining than ours, albeit able to detect sources with much higher deformations. 
}
\end{figure}

\begin{figure}[htbp]
\includegraphics[width=3.3in]{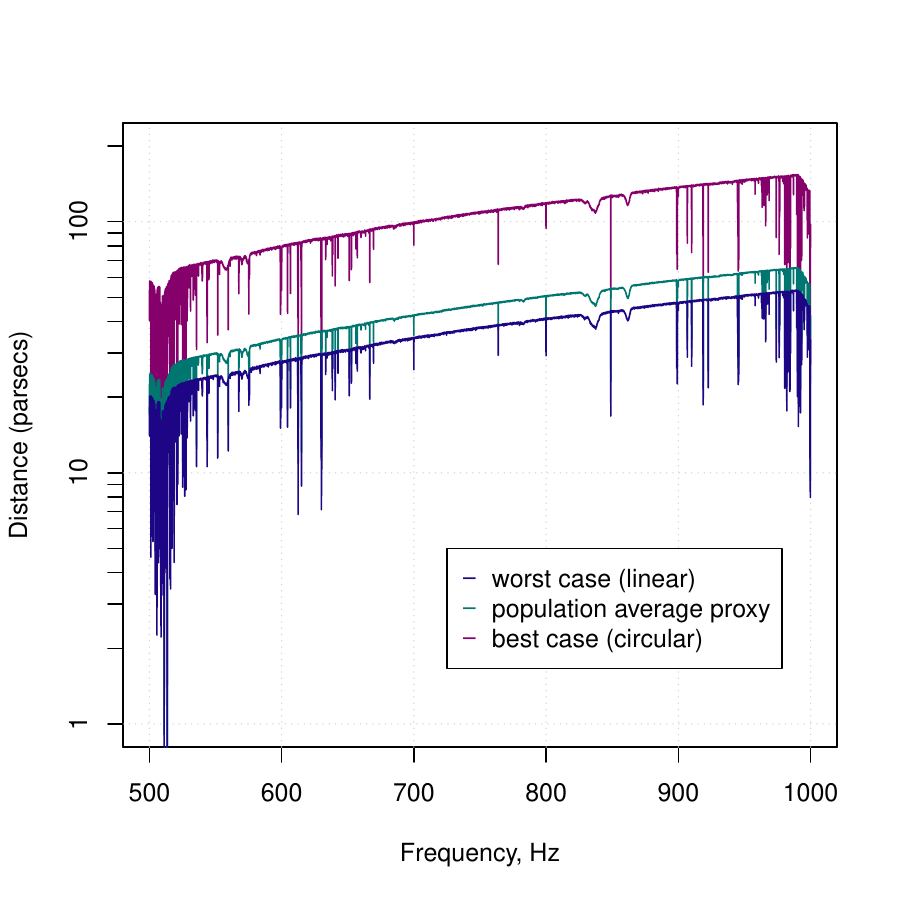}
\caption[Spindown range]{
\label{fig:distance}
Reach of the search for stars with ellipticity of $10^{-8}$. The search is also sensitive to sources with ellipticities of $10^{-7}$ with a distance from Earth that is 10 times higher. The X axis is the gravitational wave frequency, which is twice the pulsar rotation frequency for emission due to an equatorial ellipticity. R-modes and other emission mechanisms give rise to emission at different frequencies \cite{bo_rmodes}. The top curve (purple) shows the reach for a population of circularly polarized sources; The middle curve (cyan) holds for a population of sources with random orientations; The bottom curve (blue) holds for linearly polarized sources. }
\end{figure}

The Falcon pipeline computes upper limits on the intrinsic amplitude $h_0$ of continuous gravitational waveforms as a function of signal frequency and source sky position. We provide upper limits \cite{data} for circularly polarized gravitational waves, which are generated by optimally oriented sources, as well as worst-case upper limits computed by maximizing over polarization. Such upper limits hold for sources with the most unfavorable orientation.

The upper limits are part of  the frequency-resolved atlas of the sky, which also includes the measured SNR as a function of signal frequency and source sky-position. 
Specifically, the atlas contains sky-resolved data for every $45$\,mHz signal-frequency band. The sky resolution increases proportionally to frequency, and at $1000$\,Hz at any location in the sky there is a grid point within a $0.26^\circ$ radius. 
For each sky pixel and frequency bin we also provide the highest SNR value and the corresponding frequency. Our simulations show that if the maximum SNR exceeds 9 at the closest sky grid point to the signal location, the signal frequency is within $0.6$\,mHz of the frequency of the maximum, with 95\% confidence.

The upper limits are summarized in Figures \ref{fig:amplitudeULs} and \ref{fig:distance}, by taking the maximum over the sky in 45 mHz frequency bands. This is the traditional way to present continuous wave upper limit results. 

The upper limits can be taken as a measure of the sensitivity of our search, and recast in terms of the equatorial ellipticity $\varepsilon $ of sources that the search is able to detect: 
\begin{equation}
\varepsilon = { \frac{c^4}{4 \pi^2 G} } { \frac{h_0 d}{ I_{zz} {f_0}^2 } },
\label{eq:epsilon}
\end{equation}
where $I_{zz}=10^{38}$\,kg\,m$^2$ is the moment of inertia of the star with respect to the principal axis aligned with the rotation axis and $d$ its distance. 

This search is sensitive to sources with ellipticity of $10^{-8}$ up to 150\,pc away, improving on our O2 results by a factor of  $\approx$ 1.5 and on the full O3 results of \cite{lvc_O3_allsky2} and \cite{LIGOScientific:2021jlr} by a factor of $\approx$ 1.7 and 1.5, respectively. 
Signals sourced by ellipticities of $10^{-7}$ would have been detected from sources within a distance of 500 pc and, depending on orientation, as large as 1.5 kpc, at the highest frequencies. For lowest frequencies the distance is up to 200 pc, and  as large as 600 pc for optimally oriented sources. 

The results in \cite{lvc_O3_allsky2} cover a much broader spin-down range than our search. They exclude neutron stars with ellipticities larger than  $10^{-6}$ within a few kiloparsecs, motivating a search like ours for lower ellipticity sources. We accomplish this with a spindown range that covers ellipticities as high as $10^{-7}$, while improving the sensitivity of the search. 

The atlas is made possible due to our rigorous analytical upper limit procedure \cite{universal_statistics}. The analytical formula is required to efficiently produce billions of individual upper limits, and the statistical rigor is a necessity to ensure that any subset of results is individually valid regardless of any peculiarities in the  distribution of underlying data. In contrast, many searches \cite{lvc_O3_allsky2,LIGOScientific:2021jlr, keith_review, EatHO3a, Covas:2022rfg} use Monte-Carlo simulations that establish upper limits over the whole sky. This method is impractical to use on many sky locations.  

The most sensitive population-average all-sky upper limits in \cite{lvc_O3_allsky2} did not use Monte-Carlo simulations, but rather an analytical formula.  While the authors provide a scaling prescription to derive upper limits at arbitrary sky positions, its validity is not obvious. In particular, it is not clear why the scaling factor derived from  {\it average} antenna beam pattern function over the population should be representative of the 95\% {\it percentile} population amplitudes. This confusion may lead to an underestimate of the upper limits. Even for the all-sky case the formula is heuristic, at best. It was derived under the assumption of independent identically distributed Gaussian noise and neglects the variation of the noise level in real data. The deviations from Gaussianity are also ignored which  precludes from establishing rigorous upper limits.

\begin{figure*}[htbp]
\includegraphics[width=\linewidth]{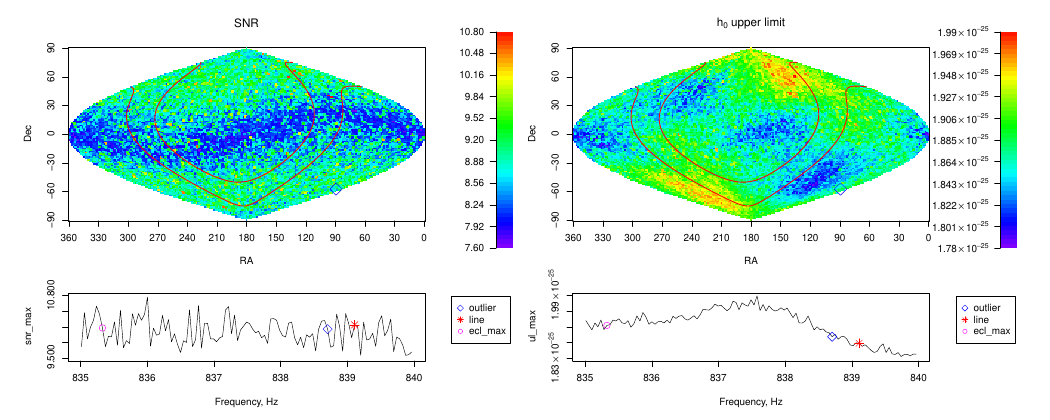}
\caption[Skymap]{
\label{fig:skymap_ul} Summary of atlas data from the bins between 835-840\,Hz. The top panels show the highest SNR (left) and upper limit values (right) across the frequency band, for each pixel of the sky map, using equatorial coordinates. The red lines denote the galactic plane.
The blue diamond shows the location of the outlier that is discarded based on the analysis of O3a+b data. The blue band of smaller SNRs near the ecliptic equator is due to large correlations between waveforms of sources in that region. The blue regions in the upper limit plot are due to the lower-SNR values in the ecliptic plane, and also occur near the ecliptic poles that are favored by the antenna pattern of the detectors. The bottom panels show the same data as a function of frequency and with the maximum taken over the sky. We mark the frequency of the band where the outlier mentioned above was found, the location of the only known line from the O3 line list in that band, and the band where we the maximum SNR is achieved in the ecliptic pole region - a region strongly affected by instrumental lines. The data and code used to produce this plot is available \cite{data}. 
}
\end{figure*}

\begin{figure*}[htbp]
\includegraphics[width=\linewidth]{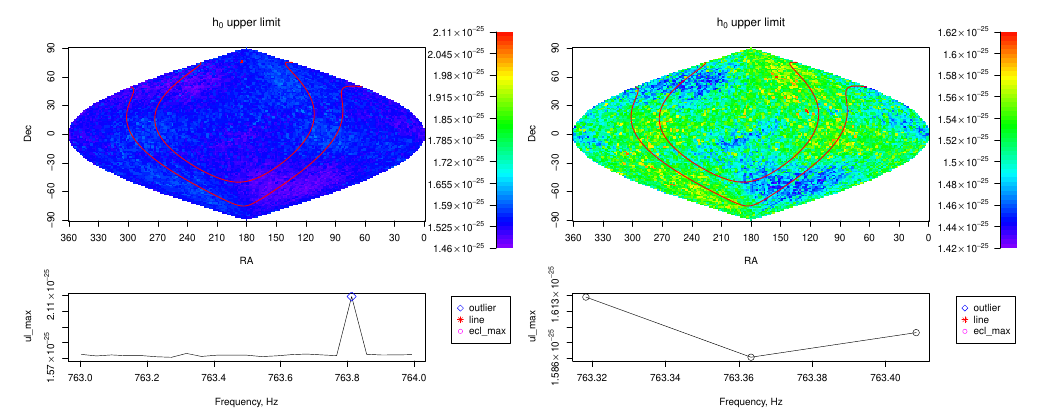}
\caption[Skymap]{
\label{fig:skymap_ul_bkg} Both panels here show upper limit results, like the right hand-side plots of Figure~\ref{fig:skymap_ul}. The frequency band is different: 763-764 Hz for the left panels and a zoom between 763.3-763.42 for the right side panels. We choose the 763-764 band to give an idea of what a signal would like like, since the hardware-injected fake signal ip9 falls in this band, see Table~\ref{tab:injections}. The signal amplitude is $1.3 \times 10^{-25}$ and it is lower than the upper limit value at the sky position, as it should be. The zoomed-in band does not contain the signal parameters and hence it is an example of a ``quiet band". }
\end{figure*}

A counterpart of our atlas is produced by the radiometer search for the gravitational wave stochastic background \cite{LIGOScientific:2021oez}. Both the radiometer and this search are affected by loud continuous signals. However, this search is specifically tuned to continuous signals as it, for example, tracks the signal Doppler shifts due to the Earth motion over months of observation. This makes it significantly more sensitive (we estimate by at least a factor of $10$) to continuous waves -- and also more computationally expensive -- than the radiometer search.

All-sky surveys for continuous gravitational waves are extremely computationally intensive: the results of this search cost in excess of 36 million CPU-hours. Our atlas \cite{data} will allow others to benefit from this investment: astronomical observations of pulsations from a certain location in the sky may be checked against high SNR occurrences at the same location and consistent frequencies; the atlas can immediately be scanned across the whole frequency range at an interesting sky position, for example that of a supernova remnant (as we demonstrate in the appendix), in search of a significant result; with an estimate of the distance of a gravitational wave source, upper limits on its ellipticity can be readily computed; models of neutron star populations can be directly convolved with the atlas upper limits to make predictions of their detectability, and the same holds for primordial black hole populations emitting continuous gravitational waves through their orbital energy loss \cite{primordial_black_holes}, or for tens of solar mass black hole populations sourcing continuous waves through super-radiance \cite{superradiance}. The fine granularity of our atlas (Figures \ref{fig:skymap_ul} and \ref{fig:skymap_ul_bkg}) produces high-number statistics and, in case of a coincidence with some other observation, this will greatly aid in the estimation of the chance probability of any finding.

\begin{table*}[htbp]
\begin{tabular}{D{.}{.}{3}D{.}{.}{5}D{.}{.}{4}D{.}{.}{4}D{.}{.}{4}l}\hline
\multicolumn{1}{c}{SNR}   &  \multicolumn{1}{c}{$f$} & \multicolumn{1}{c}{$\dot{f}$} &  \multicolumn{1}{c}{$\RAJ$}  & \multicolumn{1}{c}{$\DECJ$} & Comment \\
\multicolumn{1}{c}{}	&  \multicolumn{1}{c}{Hz}	&  \multicolumn{1}{c}{pHz/s} & \multicolumn{1}{c}{degrees} & \multicolumn{1}{c}{degrees} & \\
\hline \hline
110.4 & 763.84732 & -0.4 & 198.886 & 75.689 & Simulated signal ip9\\
98.9 & 575.16351 & -0.4 & 215.255 & 3.441 & Simulated signal ip2\\
16.7 & 993.38180 & 7.0 & 282.198 & -57.533 & Strong narrowband disturbance in H1\\
16.3 & 848.91931 & -31.6 & 39.087 & -15.706 & Simulated signal ip1\\
16.1 & 838.70499 & 1.2 & 10.883 & -57.581& Fails O3a+b consistency check\\
\hline
\end{tabular}
\caption[Outliers produced by the detection pipeline]{Outliers surviving the detection pipeline. 
Outliers marked ``ipX'' are due to simulated signals ``hardware-injected'' during the science run for validation purposes. Their parameters are listed in Table \ref{tab:injections}.  Signal frequencies refer to GPS epoch $1246070000$ (2019 Jul 2 02:33:02 UTC).} 
\label{tab:Outliers}
\end{table*}

\begin{table}[htbp]
\begin{center}
\begin{tabular}{l c c c c c }
\hline
Label & \multicolumn{1}{c}{$h_0$} &\multicolumn{1}{c}{$f$} & \multicolumn{1}{c}{$\dot{f}$} & \multicolumn{1}{c}{$\RAJ$} & \multicolumn{1}{c}{$\DECJ$} \\
 & $10^{-25}$ & \multicolumn{1}{c}{Hz} & \multicolumn{1}{c}{pHz/s} & \multicolumn{1}{c}{degrees} & \multicolumn{1}{c}{degrees} \\
\hline \hline
ip1   & 5.5 & 848.935  & -300   &   37.39     &  -29.45 \\
ip2   & 0.76 & 575.164  & -0.137   &  215.26     &    3.44 \\
ip9   & 1.3 & 763.847 & $\sci{-1.45}{-5}$    &  198.89     &   75.69 \\
\hline
\end{tabular}
\caption[Parameters of hardware injections]{Parameters of the hardware-injected simulated continuous wave signals during the O3 data run -- epoch GPS $1246070000$ (2019 Jul 2 02:33:02 UTC) -- , that fall in the parameter range probed by our search.}
\label{tab:injections}
\end{center}
\end{table}

Five outliers survive all stages (Table \ref{tab:Outliers}). Three correspond to simulated signals  \cite{hardware_injections} injected via radiation pressure on the detectors' test masses (Table \ref{tab:injections}). 

One is induced by a large instrumental artifact in H1 at 993.300\,Hz. This instrumental artifact is quite loud, but is not included in the known line list \cite{O3aDataSet}. The reanalysis  using H1 data alone yields an SNR of 16.3, while more sensitive L1 data gives an SNR of only 9.2. 

One outlier with SNR just above threshold does not correspond to any identifiable detector artifact, however a reanalysis of this outlier using public LIGO data from the O3a and O3b \cite{O3bDataSet} runs combined, yields a decrease in the SNR -- from 16 on O3a data to 13 using O3a+b -- with a 6 day coherence length. This is not consistent with what is expected from the IT2 continuous signal model (Eq.~ \ref{eq:freqEvolution}) assumed in this search.
Thus the outlier is either an astrophysical signal that deviates significatly from the IT2 model, or the outlier is due to terrestrial contamination.

\begin{acknowledgments}
The authors thank the scientists, engineers and technicians of LIGO, whose hard work and dedication produced the data that made this search possible.

The search was performed on the ATLAS cluster at AEI Hannover. We thank Bruce Allen, Carsten Aulbert and Henning Fehrmann for their support.

We are grateful to Heinz-Bernd Eggenstein for helpful feedback on the atlas.

This research has made use of data or software obtained from the Gravitational Wave Open Science Center (gw-openscience.org), a service of LIGO Laboratory, the LIGO Scientific Collaboration, the Virgo Collaboration, and KAGRA. 
\end{acknowledgments}

\appendix*

\section{Using the Falcon atlas to search supernova remnants}
\label{VelaJr}
\begin{figure}[htbp]
\includegraphics[width=3.3in]{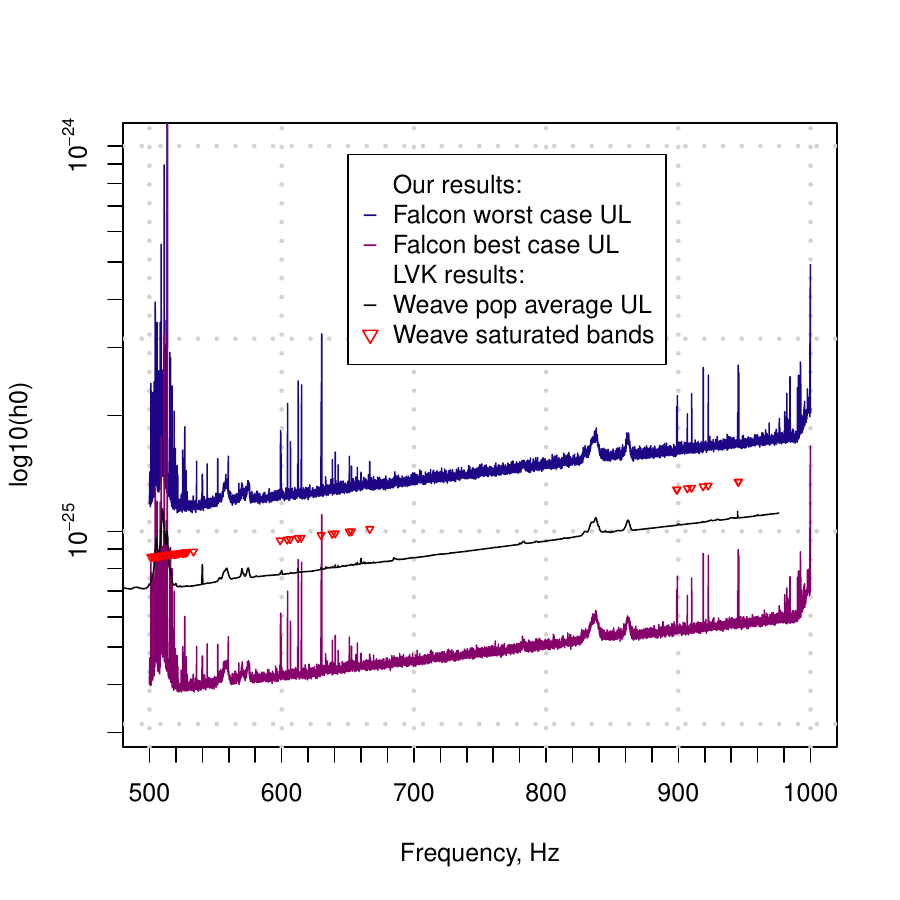}
\caption[VelaJr]{
\label{fig:velajr}
This plot shows upper limits similar to those shown in Figure \ref{fig:amplitudeULs}, but for the location of Vela Jr. Latest LIGO/Virgo/KAGRA results \cite{LVK_VelaJr} are shown for comparison. The triangles mark saturated bands for which the Weave results are invalid.
}
\end{figure}

\begin{figure}[htbp]
\includegraphics[width=3.3in]{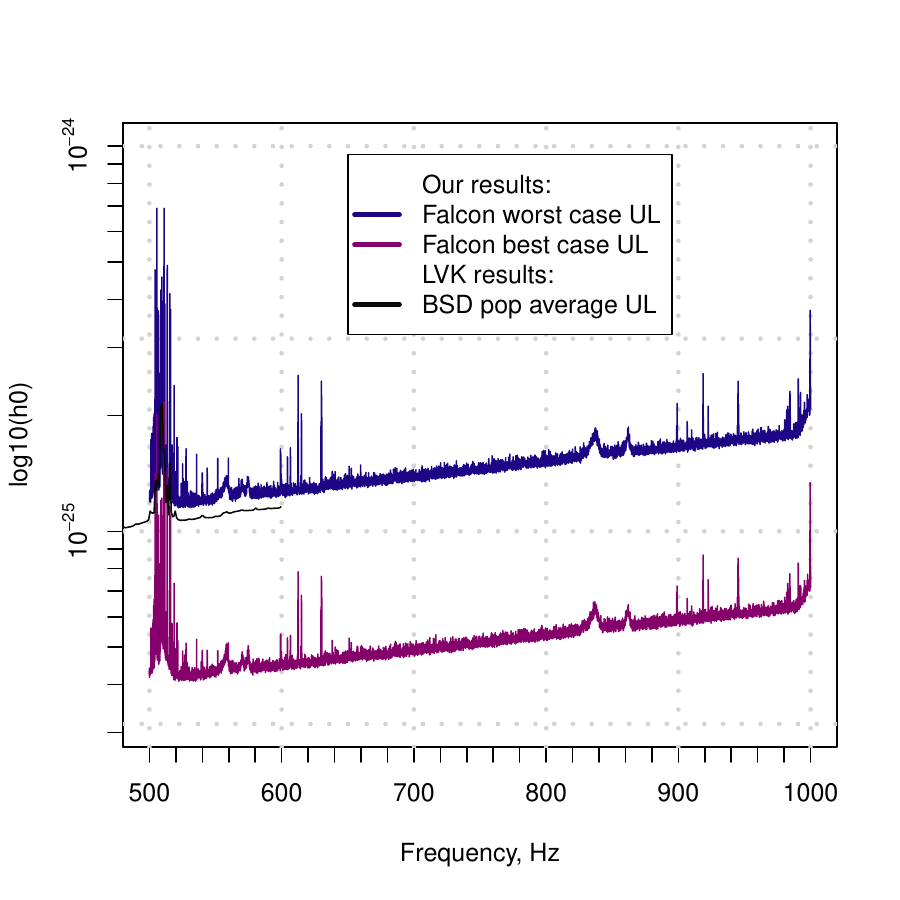}
\caption[G189]{
\label{fig:G189}
This plot shows upper limits similar to those shown in Figure \ref{fig:amplitudeULs}, but for the location of G189.1+3.0. Latest LIGO/Virgo/KAGRA results \cite{LVK_G189} are shown for comparison. The LVK upper limit curve was computed as minimum of Hanford, Livingston and Virgo data. 
}
\end{figure}

It is easy to use our atlas to examine arbitrary locations on the sky. An example R script {\tt spatial\_index\_example2.R} included with the atlas data shows how to produce upper limits and SNR values for any given sky location. By simply setting right ascension and declination to the coordinates of the Vela Jr or G189.1+3.0 supernova remnants, we obtain the data shown in Figures \ref{fig:velajr} and \ref{fig:G189} respectively. 

We also plot the latest LIGO/Virgo/KAGRA results \cite{LVK_VelaJr, LVK_G189}. Our spindown range is smaller than that searched by \cite{LVK_VelaJr, LVK_G189}, but we have no excluded bands. Furthermore, our results are rigorous 95\% confidence level upper limits, rather than sensitivity estimates.

These plots illustrate that an atlas user can establish rigorous upper limits that  improve on the best dedicated directed searches.

This data can be used as is to place limits on the gravitatational radiation coming from direction of these supernova remnants as we have done, or it can be the starting point for a deeper search.

\newpage
\begin{thebibliography}{99}

\def\etal{{\it et al.}}

\bibitem{LIGOScientific:2021djp}
R.~Abbott \textit{et al.} [LIGO Scientific, VIRGO and KAGRA],
GWTC-3: Compact Binary Coalescences Observed by LIGO and Virgo During the Second Part of the Third Observing Run,
arXiv:2111.03606

\bibitem{aligo} J.~Aasi \etal\ (LIGO Scientific Collaboration), Advanced LIGO,  Class.\ Quantum Grav.\  {\bf 32} 7 (2015) 

\bibitem{avirgo} F.~Acernese \etal, Advanced Virgo: a second-generation interferometric gravitational wave detector, Class.\ Quantum Grav.\ { \bf 32} 024001 (2015)

\bibitem{KAGRA:2022fgc}
H.~Abe \textit{et al.} [KAGRA],
Performance of the KAGRA detector during the first joint observation with GEO 600 (O3GK), Progress of Theoretical and Experimental Physics, ptac093 (2022)
arXiv:2203.07011

\bibitem{Haskell:2015psa}
B.~Haskell, M.~Priymak, A.~Patruno, M.~Oppenoorth, A.~Melatos and P.~D.~Lasky,
Detecting gravitational waves from mountains on neutron stars in the Advanced Detector Era,
Mon. Not. Roy. Astron. Soc. \textbf{450}, no.3, 2393-2403 (2015)



\bibitem{Lindblom:1998wf}
L.~Lindblom, B.~J.~Owen and S.~M.~Morsink, Gravitational radiation instability in hot young neutron stars,
Phys. Rev. Lett. \textbf{80}, 4843-4846 (1998)

\bibitem{bo_rmodes}
B.J.~Owen, How to adapt broad-band gravitational-wave searches for r-modes, Phys. Rev. D \textbf{82} 104002 (2010)

\bibitem{superradiance}
A.~Arvanitaki and S.~Dubovsky, Exploring the string axiverse with precision black hole physics, Phys. Rev. D  {\bf 83} 044026 (2011) 

\bibitem{LIGOScientific:2021jlr}
R.~Abbott \textit{et al.} (LIGO Scientific, VIRGO and KAGRA collaborations),
All-sky search for gravitational wave emission from scalar boson clouds around spinning black holes in LIGO O3 data,
arXiv:2111.15507

\bibitem{Zhu:2020tht}
S.~J.~Zhu, M.~Baryakhtar, M.~A.~Papa, D.~Tsuna, N.~Kawanaka and H.~B.~Eggenstein,
Characterizing the continuous gravitational-wave signal from boson clouds around Galactic isolated black holes,
Phys. Rev. D \textbf{102}, no.6, 063020 (2020)

\bibitem{Arvanitaki:2009fg}
A.~Arvanitaki, S.~Dimopoulos, S.~Dubovsky, N.~Kaloper and J.~March-Russell,
Phys. Rev. D \textbf{81}, 123530 (2010)
doi:10.1103/PhysRevD.81.123530

\bibitem{keith_review} K.~Riles,
Searches for Continuous-Wave Gravitational Radiation,
arXiv:2206.06447

\bibitem{loosely_coherent} 
V.~Dergachev, On blind searches for noise dominated signals: a loosely coherent approach, Class.\ Quantum Grav.\ {\bf 27}, 205017 (2010).

\bibitem{loosely_coherent2} 
V.~Dergachev, Loosely coherent searches for sets of well-modeled signals,
Phys.\ Rev.\ D {\bf 85}, 062003 (2012)  

\bibitem{loosely_coherent3} 
V.~Dergachev, Loosely coherent searches for medium scale coherence lengths, arXiv:1807.02351   

\bibitem{O2_falcon}
V.~Dergachev, M.~A.~Papa, 
Results from the first all-sky search for continuous gravitational waves from small-ellipticity sources, 
Phys. Rev. Lett. \textbf{125}, no.17, 171101  (2020)

\bibitem{O2_falcon2}
V.~Dergachev, M.~A.~Papa, 
Results from high-frequency all-sky search for continuous gravitational waves from small-ellipticity sources,  Phys. Rev. {\bf D} 103, 063019 (2021)


\bibitem{O2_falcon3}
V.~Dergachev, M.~A.~Papa, 
The search for continuous gravitational waves from small-ellipticity sources at low frequencies, Phys. Rev. {\bf D} 104, 043003 (2021)

\bibitem{EatHO3a} B.~Steltner, M.A.~Papa, H.-B.~Eggenstein, B.~Allen, V.~Dergachev,  B.~Machenschalk, O.~Behnke, O. and R.~Prix, Deep Einstein@Home all-sky search for continuous gravitational waves in LIGO O3 public data, arXiv:2303.04109  (2023)


\bibitem{JST}
Jonathan P. Gardner, John C. Mather, Mark Clampin, Rene Doyon, Matthew A. Greenhouse, Heidi B. Hammel, John B. Hutchings, Peter Jakobsen, Simon J. Lilly, Knox S. Long, Jonathan I. Lunine, Mark J. McCaughrean, Matt Mountain, John Nella, George H. Rieke, Marcia J. Rieke, Hans-Walter Rix, Eric P. Smith, George Sonneborn, Massimo Stiavelli, H. S. Stockman, Rogier A. Windhorst, Gillian S. Wright,
The James Webb Space Telescope,
Space Sci.Rev. 123 (2006) 485

\bibitem{CTA}
B.S.~Acharya \etal (Cherenkov Telescope Array Consortium), 
Science with the Cherenkov Telescope Array,
arXiv:1709.07997

\bibitem{SKA}
P. E. Dewdney, P. J. Hall, R. T. Schilizzi and T. J. L. W. Lazio, "The Square Kilometre Array," in Proceedings of the IEEE, vol. 97, no. 8, pp. 1482-1496, Aug. 2009, 

\bibitem{AMS}
M.~Aguilar \etal,
{The Alpha Magnetic Spectrometer (AMS) on the international space station: Part II \textemdash{} Results from the first seven years},
Phys.Rept. 894 (2021) 1-116

\bibitem{PTF}
N.M.~Law \etal,
The Palomar Transient Factory: System Overview, Performance and First Results,
arXiv:0906.5350 

\bibitem{ATNF}
R.N.~Manchester, G.B.~Hobbs, A.~Teoh, M.~Hobbs, 
The Australia Telescope National Facility Pulsar Catalogue,
Astron. J., 129, 1993-2006 (2005) 

\bibitem{SIMBAD}
M.  Wenger, F.  Ochsenbein, D.  Egret, P.  Dubois, F.  Bonnarel, S.  Borde, F.  Genova, G.  Jasniewicz, S.  Laloë, S.  Lesteven, R.  Monier,
The SIMBAD astronomical database - The CDS reference database for astronomical objects,
Astron. Astrophys. Suppl. Ser. 143 (1) 9-22 (2000)

\bibitem{MVL} Mappable Vector Library,
\url{https://github.com/volodya31415/libMVL}

\bibitem{RMVL}
R package for Mappable Vector Library
\url{https://cran.r-project.org/package=RMVL}


\bibitem{Johnson-McDaniel:2012wbj}
N.K.~Johnson-McDaniel and B.J.~Owen,
Maximum elastic deformations of relativistic stars,
Phys. Rev. D \textbf{88}, 044004 (2013)

\bibitem{Gittins:2020cvx}
F.~Gittins, N.~Andersson and D.~I.~Jones,
Modelling neutron star mountains,
Mon. Not. Roy. Astron. Soc. \textbf{500}, no.4, 5570-5582 (2020)

\bibitem{Gittins:2021zpv}
F.~Gittins and N.~Andersson,
Modelling neutron star mountains in relativity,
Mon. Not. Roy. Astron. Soc. \textbf{507}, no.1, 116-128 (2021)

\bibitem{Pagliaro:2023bvi}
G.~Pagliaro, M.~A.~Papa, J.~Ming, J.~Lian, D.~Tsuna, C.~Maraston and D.~Thomas,
Continuous gravitational waves from Galactic neutron stars: demography, detectability and prospects, 
arXiv:2303.04714 [gr-qc].

\bibitem{ellipticity}
G.~Woan, M.~D.~Pitkin, B.~Haskell, D.~I.~Jones, and P.~D.~Lasky, Evidence for a Minimum Ellipticity in Millisecond Pulsars, ApJL {\bf{863}} L40 (2018)

\bibitem{Jaranowski:1998qm}
P.~Jaranowski, A.~Krolak and B.~F.~Schutz,
Data analysis of gravitational - wave signals from spinning neutron stars. 1. The Signal and its detection,
Phys. Rev. D \textbf{58} (1998), 063001


\bibitem{Fourier}S.{\hspace{0.167em}
}R. Valluri and V. Dergachev and X. Zhang and F.{\hspace{0.167em}}A. Chishtie, 
Fourier transform of the continuous gravitational wave signal, 
Phys. Rev. D \textbf{104}, no.2, 024065 (2021)

\bibitem{universal_statistics}
V.~Dergachev, A Novel Universal Statistic for Computing Upper Limits in Ill-behaved Background,
Phys.\ Rev.\ D \textbf{87}, 062001 (2013).

\bibitem{ATLAS}
ATLAS computing cluster
\url{https://www.aei.mpg.de/43564/atlas-computing-cluster}


\bibitem{O3aDataSet} The O3a Data Release
 \url{https://doi.org/10.7935/nfnt-hm34}
 

\bibitem{O3bDataSet} The O3b Data Release
\url{https://doi.org/10.7935/pr1e-j706}


\bibitem{Papa:2020vfz}
M.~A.~Papa, J.~Ming, E.~V.~Gotthelf, B.~Allen, R.~Prix, V.~Dergachev, H.~B.~Eggenstein, A.~Singh and S.~J.~Zhu,
Search for Continuous Gravitational Waves from the Central Compact Objects in Supernova Remnants Cassiopeia A, Vela Jr., and G347.3\textendash{}0.5,
Astrophys. J. \textbf{897}, no.1, 22 (2020)

\bibitem{lvc_O3_allsky}
R.~Abbott \etal (LIGO Scientific Collaboration, Virgo Collaboration and KAGRA Collaboration), All-sky search for continuous gravitational waves from isolated neutron stars in the early O3 LIGO data, Phys. Rev. {\bf D} 104, 082004 (2021) 

\bibitem{O2O3_detchar}
D.~Davis \etal, LIGO detector characterization in the second and third observing runs, Class. Quantum Grav. {\bf 38} 135014 (2021)

\bibitem{Steltner:2021qjy}
B.~Steltner, M.~A.~Papa and H.~B.~Eggenstein,
Identification and removal of non-Gaussian noise transients for gravitational-wave searches,
Phys. Rev. D \textbf{105}, no.2, 022005 (2022)
doi:10.1103/PhysRevD.105.022005

\bibitem{data} See EPAPS Document No. [number will be inserted by publisher] for numerical values of upper limits. Also at \url{https://www.aei.mpg.de/continuouswaves/O3aFalcon500-1000} including full atlas.
 
\bibitem{lvc_O3_allsky2}
R.~Abbott \etal (LIGO Scientific Collaboration, Virgo Collaboration and KAGRA Collaboration), All-sky search for continuous gravitational waves from isolated neutron stars using Advanced LIGO and Advanced Virgo O3 data, arXiv:2201.00697




\bibitem{Covas:2022rfg}
P.~B.~Covas, M.~A.~Papa, R.~Prix and B.~J.~Owen,
Constraints on r-modes and Mountains on Millisecond Neutron Stars in Binary Systems,
Astrophys. J. Lett. \textbf{929}, no.2, L19 (2022)


\bibitem{LIGOScientific:2021oez}
R.~Abbott \etal (LIGO Scientific Collaboration, Virgo Collaboration and KAGRA Collaboration),
All-sky, all-frequency directional search for persistent gravitational-waves from Advanced LIGO's and Advanced Virgo's first three observing runs,
arXiv:2110.09834.

\bibitem{primordial_black_holes}
A.L.~Miller, S.~Clesse, F.~De~Lillo, G.~Bruno, A.~Depasse, A.~Tanasijczuk,
Physics of the Dark Universe 32 100836 (2021)

\bibitem{hardware_injections} 
C.~Biwer, D.~Barker, J.C.~Batch, J.~Betzwieser, R.P.~Fisher, E.~Goetz, \etal, Validating gravitational-wave detections: The Advanced LIGO hardware injection system,  Phys. Rev. D {\bf 95}, 062002 (2017)

\bibitem{LVK_VelaJr}
R.~Abbott \etal (LIGO Scientific Collaboration and Virgo),
Search of the early O3 LIGO data for continuous gravitational waves from the Cassiopeia A and Vela Jr. supernova remnants, Phys. Rev. D {\bf 105}, 082005 (2022)

\bibitem{LVK_G189}
R.~Abbott \etal (LIGO Scientific Collaboration, Virgo Collaboration and KAGRA Collaboration),
Searches for Continuous Gravitational Waves from Young Supernova Remnants in the Early Third Observing Run of Advanced LIGO and Virgo,  ApJ 921 80 (2021)


\end{thebibliography}
\end{document}